\pdfoutput=1
\documentclass[namedreferences]{SolarPhysics}
\usepackage[optionalrh]{spr-sola-addons} 
\usepackage{graphicx}                    
\usepackage{color}                       
\usepackage{url}                         
\usepackage{lscape}
\usepackage{rotating}
\usepackage[pdfborder={0 0 0 },urlcolor=blue,breaklinks]{hyperref}

\begin{document}

\begin{article}

\begin{opening}

\title{Apparent Solar Tornado\,-\,Like Prominences}

\author{Olga~\surname{Panasenco}$^{1}$
\sep Sara F.~\surname{Martin}$^{2}$ 
\sep Marco~\surname{Velli}$^{3}$
}

%
\runningauthor{O. Panasenco \it{et al.}}
\runningtitle{Tornado\,-\,Like Prominences}


\institute{$^1$ Advanced Heliophysics, Pasadena, CA, USA \\ 
email: \url{panasenco.olga@gmail.com} \\
$^2$ Helio Research, La Crescenta, CA, USA \\
$^3$ Jet Propulsion Laboratory, California Institute of Technology, Pasadena, CA, USA\\
}

\begin{abstract}
Recent high-resolution observations from the \emph{Solar Dynamics Observatory} (SDO) have reawakened interest in the old and fascinating phenomenon of solar tornado-like prominences. This class of prominences was first introduced by Pettit (1932), who studied them over many years. Observations of tornado prominences similar to the ones seen by SDO had already been documented by Secchi (1877) in his famous \emph{Le Soleil}. High resolution and high cadence multiwavelength data obtained by SDO reveal that the tornado-like appearance of these prominences is mainly an illusion due to projection effects. We discuss two different cases where prominences on the limb might appear to have a tornado-like behavior.   One case of apparent vortical motions in prominence spines and barbs arises from the (mostly) 2D counterstreaming plasma motion along the prominence spine and barbs together with oscillations along individual threads. The other case of apparent rotational motion is observed in prominence cavities and results from the 3D plasma motion along the writhed magnetic fields inside and along the prominence cavity as seen projected on the limb. Thus, the ``tornado" impression results either from counterstreaming and oscillations or  from the projection on the plane of the sky of plasma motion along magnetic field lines, rather than from a true vortical motion around an (apparent) vertical or horizontal axis. We discuss the link between tornado-like prominences, filament barbs, and photospheric vortices at their base.  
\end{abstract}


\end{opening}

\section{Introduction}

Prominences closely resembling terrestrial tornadoes in form, when projected on the solar limb, have been observed spectroscopically since 1868. Probably the first published drawings, similar to the tornado-like prominences recently observed by the {\it Transition Region and Coronal Explorer} (TRACE: Handy {\it et al.}, 1999) and the {\it Atmospheric Imaging Assembly} (AIA: Lemen {\it et al}., 2012) onboard the {\it Solar Dynamics Observatory} (SDO), can be found in Secchi (1877) under the type ``Flammes''. Young (1896) described these transient structures as resembling ``whirling waterspouts, capped by a great cloud". The first successful photographs of tornado-like prominences, made by Slocum in 1910 and by Pettit in 1919, were published by Pettit (1925). Pettit preserved his interest in this subject and periodically came back to its study (Pettit, 1932; 1941; 1943; 1946; 1950). Pettit summarized the appearance of tornado-like prominences as "Vertical spirals or tightly twisted ropes" (Pettit, 1932), and ``In silhouette, tornado prominences are É columnar, usually with a small smoke-like streamer issuing from the top, often bent over, even touching the chromosphere" (Pettit, 1950). Most prominences of this type are from 5000 to 22\,000 km in width and 25\,000 to 100\,000 km in height (Pettit, 1943). The tornado-like prominences described so far are transient objects, often appearing in groups. 

\begin{figure}
\center
\includegraphics[scale=.14]{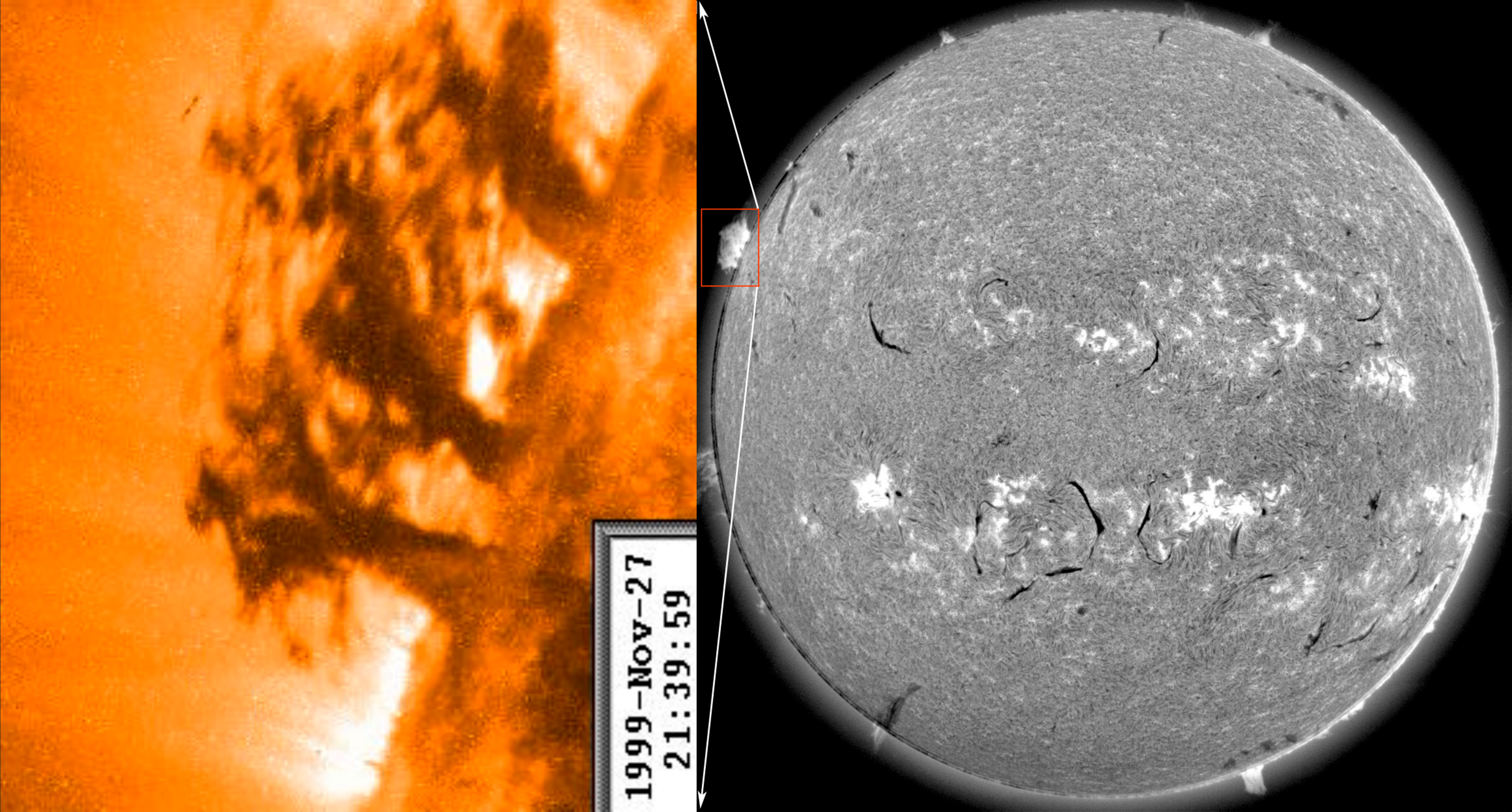}
\caption{A group of tornado-like prominences observed by TRACE in the 171\,\AA\, spectral line on 27 November 1999 (left).  The corresponding prominence structure on the limb as seen in a full disk H$\alpha$ image from BBSO (right). The orange box around the prominence in the full disk image corresponds to the TRACE field of view and image in the left panel.}
\end{figure}

Recent high-resolution observations from SDO have reawakened interest in apparent solar tornado-like prominences.  The higher spatial resolution and greater temporal cadence of the SDO observations together with their unprecedented duration and continuity shed new light on the possible origin, formation, and evolution of solar tornado-like prominences. Recent works in this area have described the SDO observations of apparent tornado-like prominences as helical structures with rotational motions (Li {\it et al}., 2012; Su {\it et al}., 2012; Orozco Su{\'a}rez {\it et al}., 2012; Panesar {\it et al}., 2013). We question this geometry and describe alternative models and mechanisms of formation. We also discuss the role that photospheric vortices may play in their dynamics (Velli and Liewer, 1999; Attie {\it et al}., 2009; Wedemeyer-B{\"o}hm {\it et al}., 2012; Rapazzo, Velli, and Einaudi, 2013; Kitiashvili {\it et al}., 2013). 

The first spacecraft observations of this phenomenon were done with TRACE (Figure 1). The TRACE movie from 27 November 1999 (see supplementary materials) seems to show a tornado-like motion at the limb, which can also be interpreted in terms of the counterstreaming of prominence plasma, with velocities up to 50 km s$^{-1}$, along the local magnetic field lines comprising the prominence spine and barbs (Zirker, Engvold, and Martin, 1998; Lin, Engvold, and Wiik, 2003; Lin {\it et al}., 2007, 2012). Both the filament spine and barbs are composed of thin threads (Lin {\it et al}., 2005a), but the barbs are threads or groups of threads that branch from the axis of the filament to the chromosphere on each side of the filament.  Barbs are not a ubiquitous feature of prominences, as there is a continuous spectrum of filaments: from smaller, short-lived active region filament with no barbs, up to huge long-lasting quiescent filaments that have very large barbs (Martin, Lin, and Engvold, 2008).

\begin{figure}
\center
\includegraphics[scale=.26]{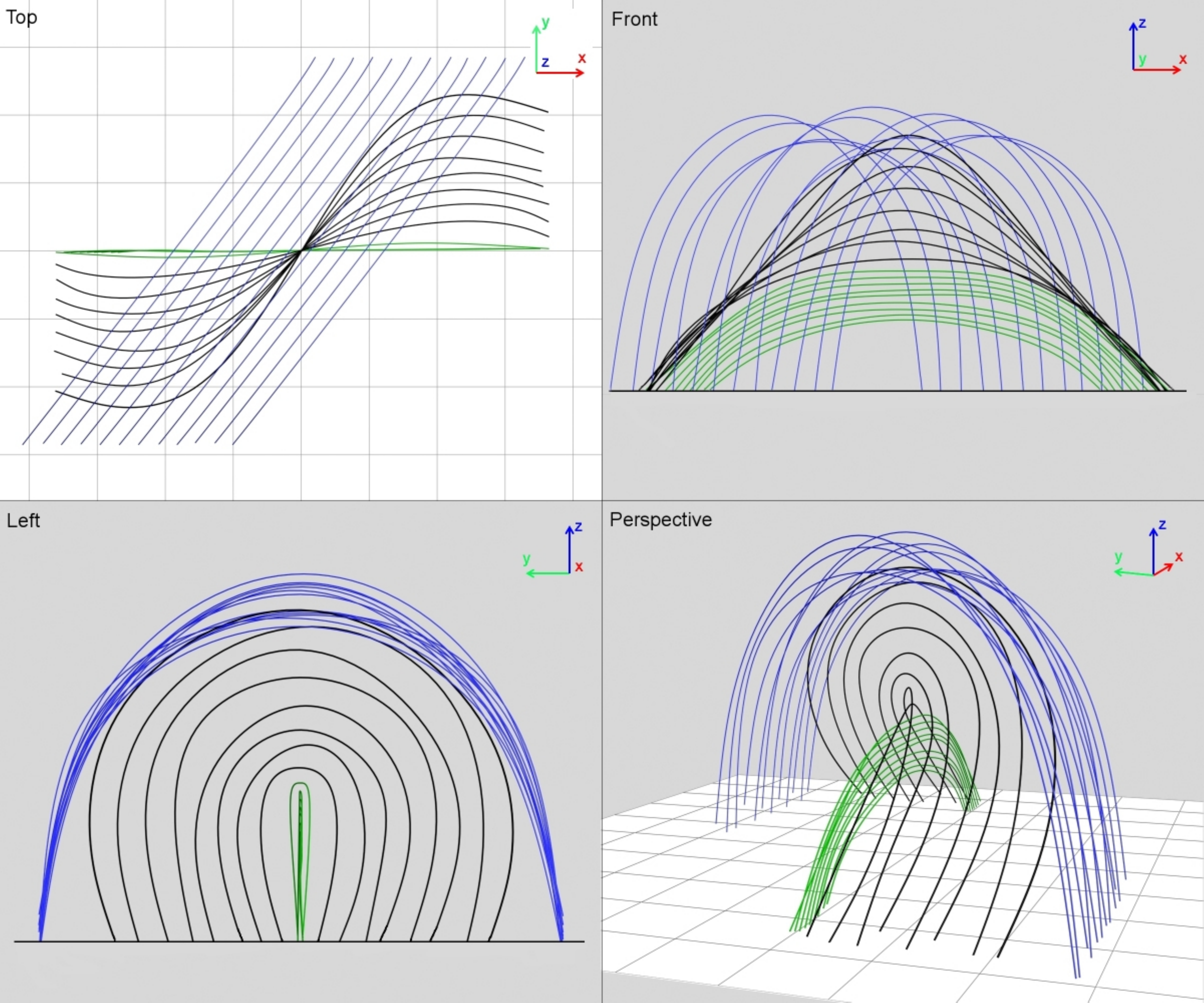}
\caption{A 3D representation of the filament channel topological system illustrating the three main components: filament spine (green), coronal loops (blue) and filament cavity (black: the space between the filament and overlying coronal loops). Cavity field lines have a strong writhe (top view), which appears as a loop in the perspective view. This writhe is the result of the interaction of the cavity field with the overlying filament arcade. The filament spine is modeled as a ribbon-like structure, and the overlying filament channel coronal arcade shows the left-skew in this model for dextral filaments (see review by Martin, 1998).}
\end{figure}

\begin{figure}
\center
\includegraphics[scale=.44]{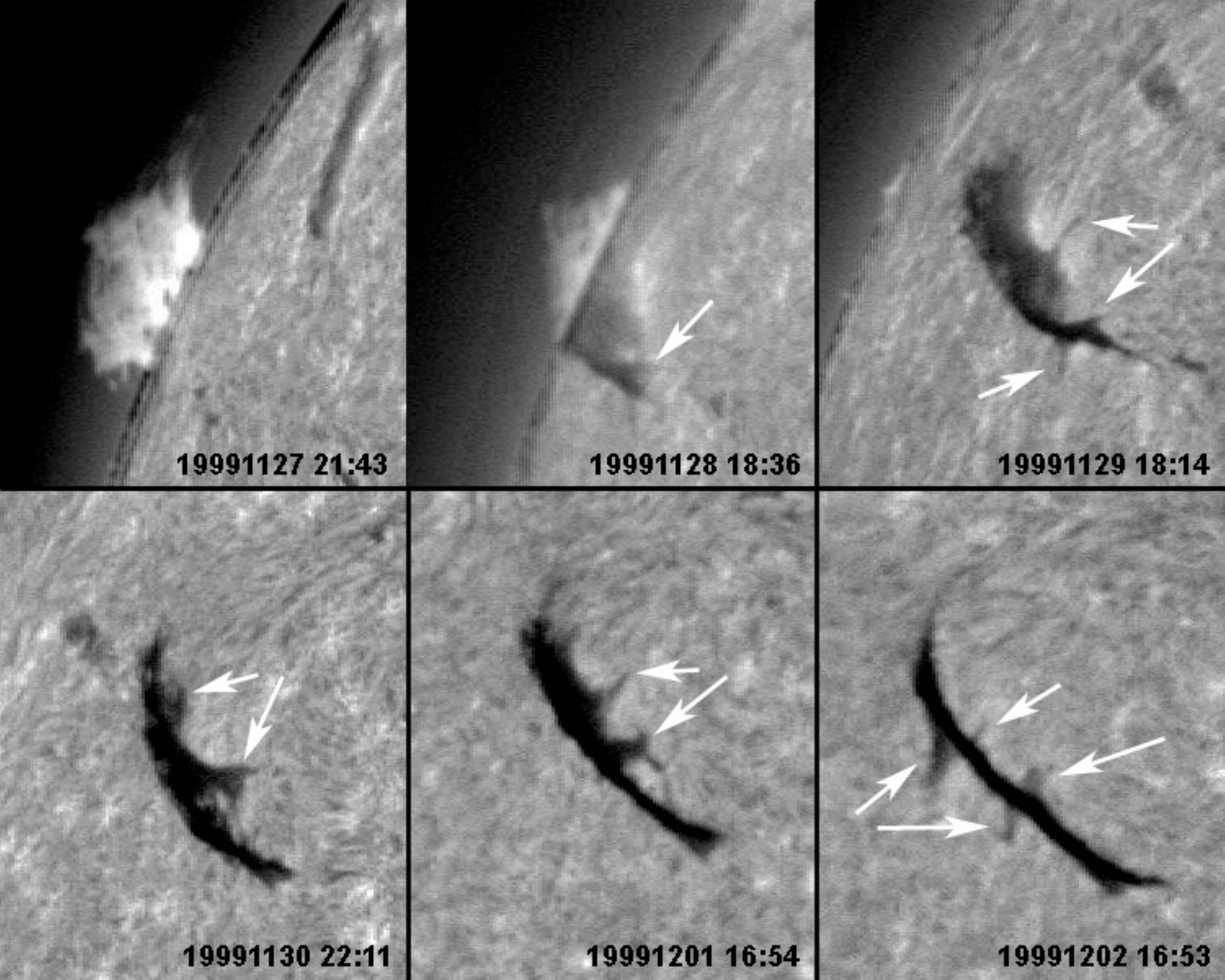}
\caption{Apparent tornado-like prominences gradually become visible against the disk -- as shown in this sequence of images from BBSO -- firstly from the side and, in the last image shown here, from the top. The vertical trunk-like parts of the ``tornados", which connect the horizontal coronal part of the limb prominence with the low chromosphere, become visible as filament barbs -- easily observed on the disk. Scales on the limb: height of the prominence $\approx$\,50\,--\,60 Mm; width of the individual ``tornado" trunk at the bottom $\approx$\,6\,-\,10 Mm; at the top $\approx$\,100 Mm. On the disk:  length of the filament $\approx$\,200 Mm, width of the filament spine $\approx$\,5\,--\,9 Mm or less; projected length of the barbs on the disk $\approx$\,40 Mm;  projected width of the barbs on the disk 5\,--\,18 Mm.}
\end{figure}
  
\begin{figure}
\center
\includegraphics[scale=.20]{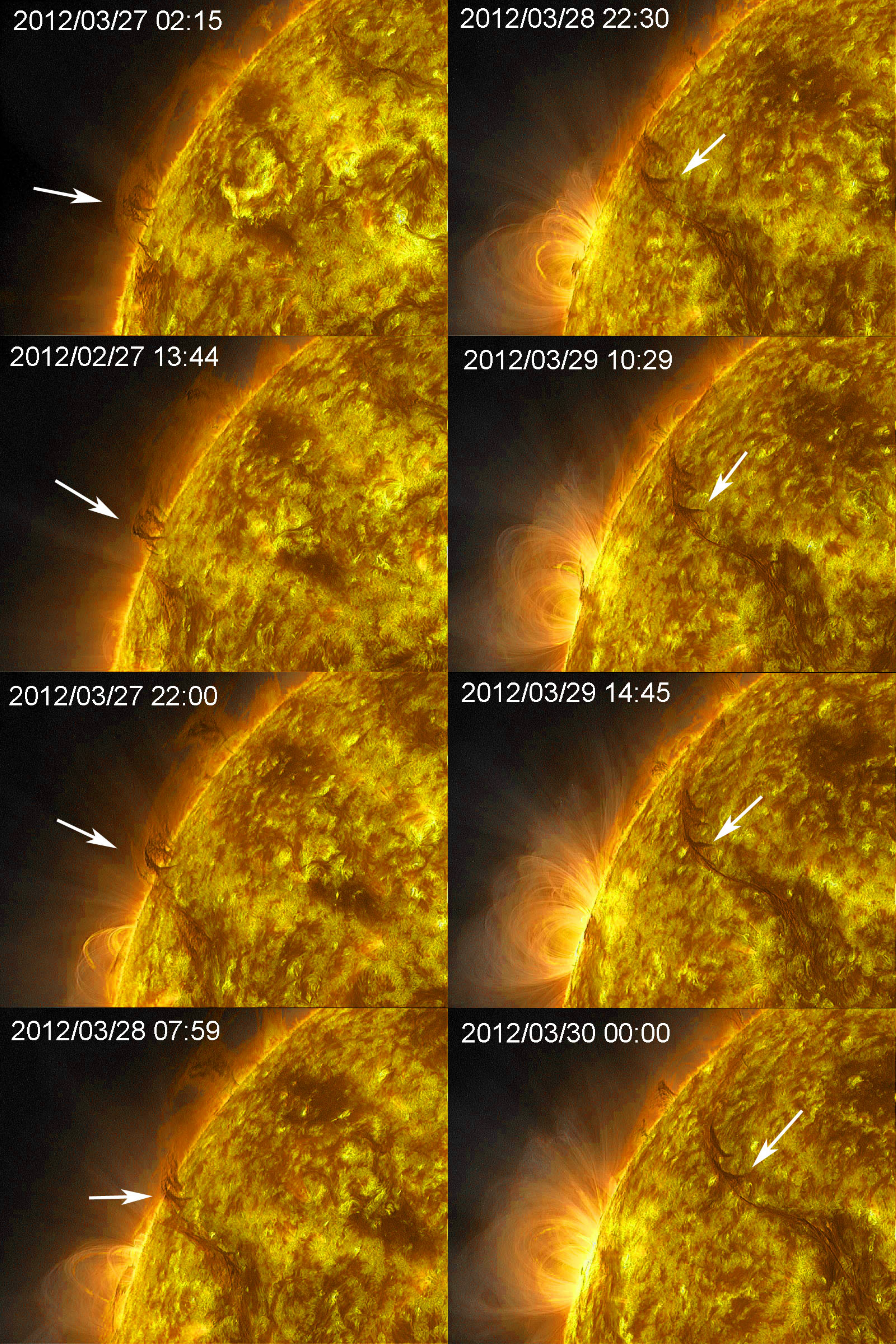}
\caption{Composite images from SDO/AIA 193\,\AA, 171\,\AA, and 304\,\AA\ spectral lines. The width of the apparent ``tornado" at the limb near its top is $\approx$ 40 Mm; the width of the filament spine at the bending point in the last image is $\approx$ 2 Mm in 193\,\AA\ and $\approx$ 3 Mm in 304\,\AA. White arrows point to the prominence/filament barb Ð the source of the tornado illusion. As the Sun rotates the apparent ``tornado" gradually becomes just a barb. }
\end{figure}
   
Barbs are threads that no longer run the full length of the spine because they have reconnected to other fields beneath or to the side of the spine. They are secondary to the spine in that they form after the spine and do not seem to prevent the spine from erupting (Martin and Echols, 1994; Martin, Bilimoria, and Tracadas, 1994; Gaizauskas {\it et al}., 1997; Engvold, 1998; Wang and Muglach, 2007; Martin, Lin, and Engvold, 2008). They readily detach from the spine during eruption, presumably by magnetic reconnection, and cease to exist (Martin and McAllister, 1997). The part of the barb above the reconnection site in the corona appears to shrink upward and merge into the expanding spine. Counterstreaming plasma is observed along the threads of the filament spine and barbs. At the solar limb, the counterstreaming plasma motion along the prominence spine and barbs will create, in specific circumstances, an illusion of rotational motion around the prominence barbs, which also have a vertical geometry and connect the mostly horizontal prominence spine to the chromosphere. We will call this limb appearance of the mostly two-dimensional motion along the prominence spine and barbs the ``vortical illusion" of apparent tornado-like prominences.

\begin{landscape}
 \begin{figure}
\center
\includegraphics[scale=.27]{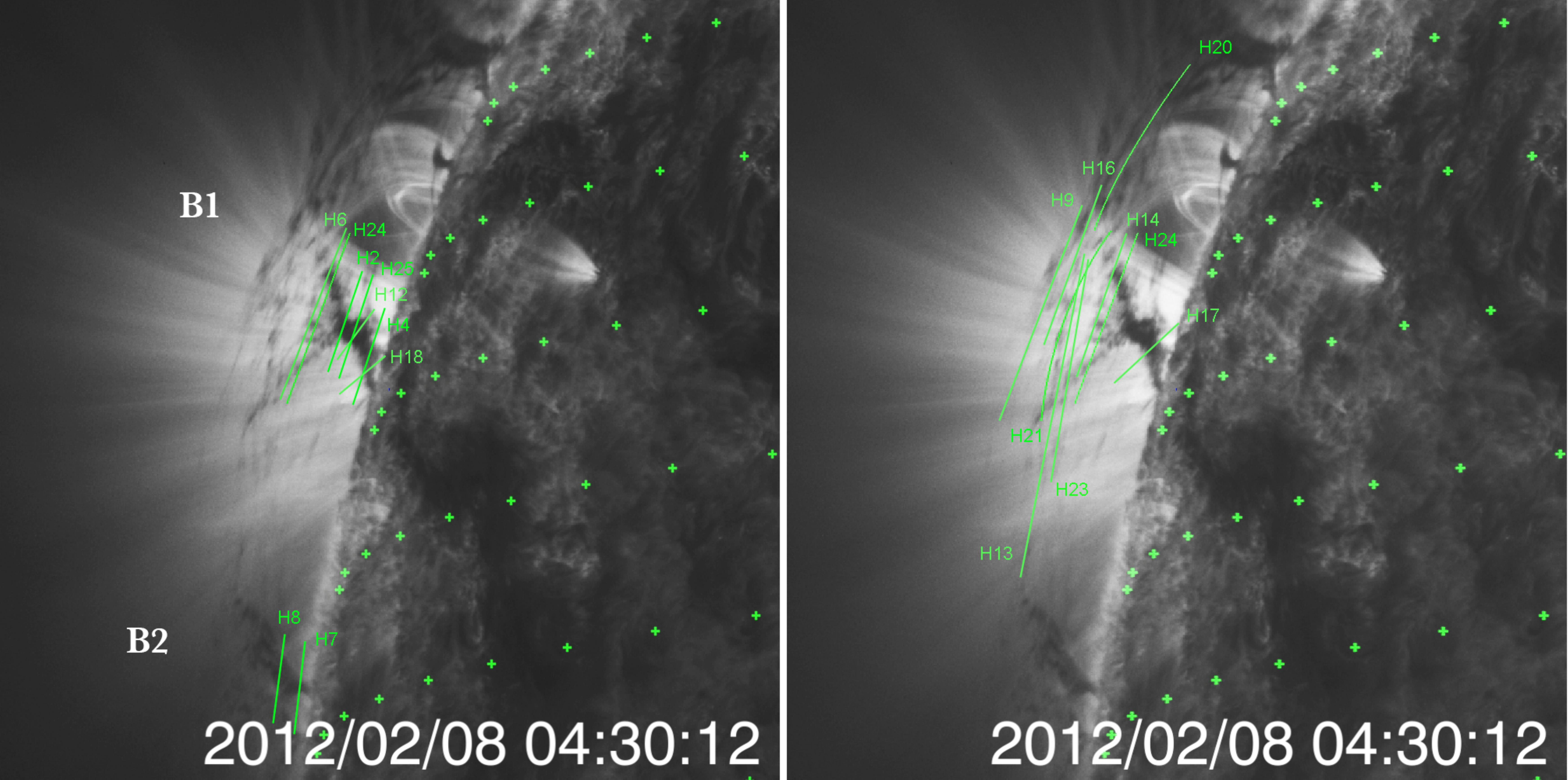}
\caption{SDO/AIA image of the prominence spine and barbs on the limb taken on 8 February 2012, in the 171\,\AA\ spectral line. The left panel shows cuts used for the subsequent time-slice diagram in the lower barb regions, while the right panel displays the cuts used for time-slice diagrams of the upper barb region and spine. Crosses mark a longitude--latitude grid with five degree separation on the Sun. The width of each cut is 3 pix $\approx$ 1.3 Mm across. Each cut has its origin at the northern point, and each cut's length is measured southward.}
\end{figure}
\end{landscape}
  
Another tornado-like motion observed only at the solar limb is caused by the sporadic plasma flow along the prominence cavity, the three-dimensional volume between the prominence spine and overlying coronal arcade. This phenomenon is a pure 2D projection on the plane of the sky of the 3D motion along the prominence cavity magnetic field lines. Implied cavity magnetic field lines generally possess writhe (see Figures 3 and 4 of T{\"o}r{\"o}k {\it et al}. (2010)). The field-line writhe gradually increases with height, starting from the top of the ribbon-shaped prominence spine and becoming very strong as one moves  upwards through the cavity cross-section to the low part of the overlying arcade (Panasenco and Martin, 2008; Martin {\it et al}., 2012). When observed from the appropriate angle, the writhed field will appear as a loop (black lines in Figure~2), and the corresponding motion along the writhed lines will be observed as a writhing   motion creating an illusion of a tornado. 
    
We now discuss observational and 3D modeling aspects of the two different plasma motions and their relationships to magnetic topology. 

\begin{figure}
\center
\includegraphics[scale=.23]{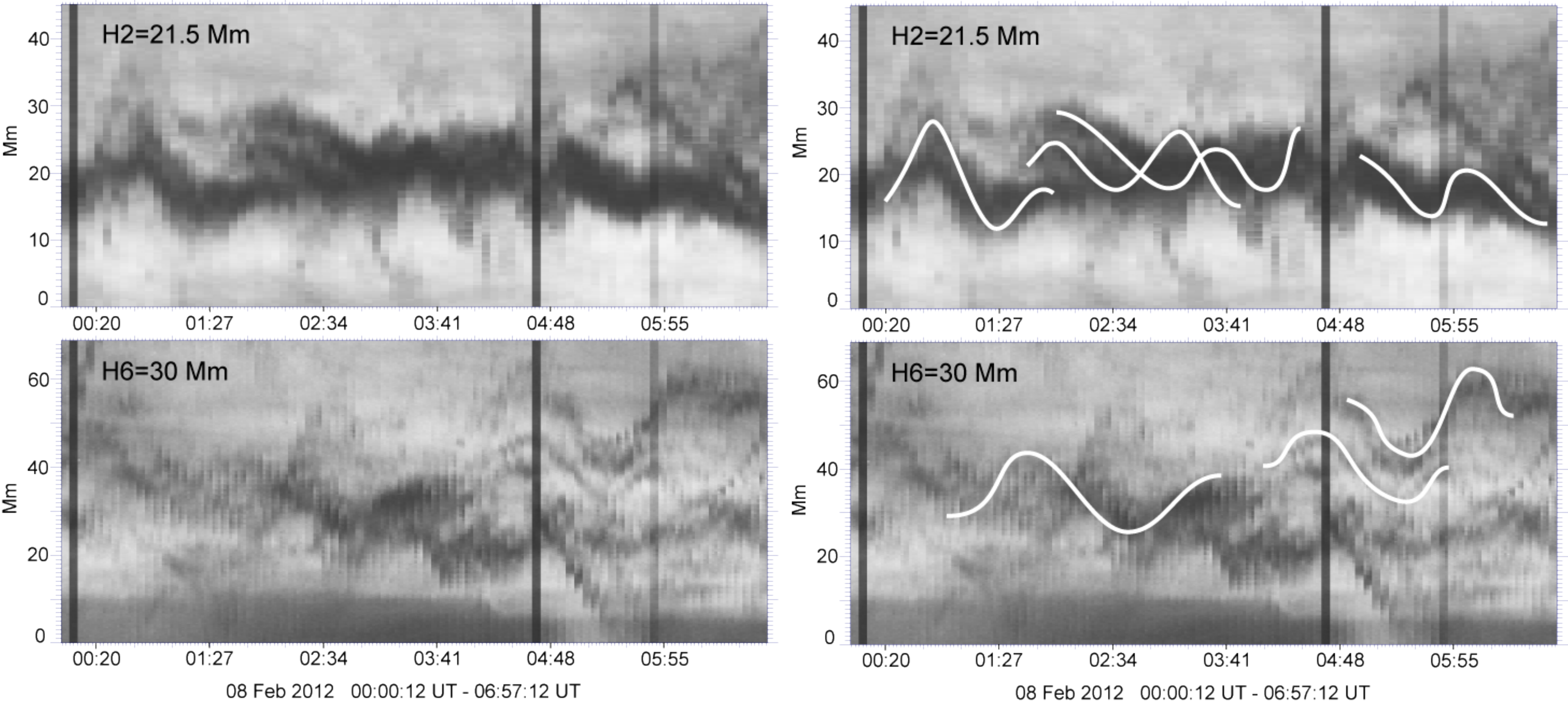}
\caption{Left panel: time-slice diagrams for cuts H2 and H6 in prominence barb B1 from Figure~5 taken by SDO/AIA in the 171\,\AA\ spectral line with five-minute cadence over $\approx$ 7 hours on 8 February 2012. Right panel: same time-slice diagrams with lines showing barb sub-structure as an aid to illustrate calculation of oscillation periodicities, ranging from 45 to 70 minutes for the sub-structures.}
\end{figure}

\begin{figure}
\center
\includegraphics[scale=.23]{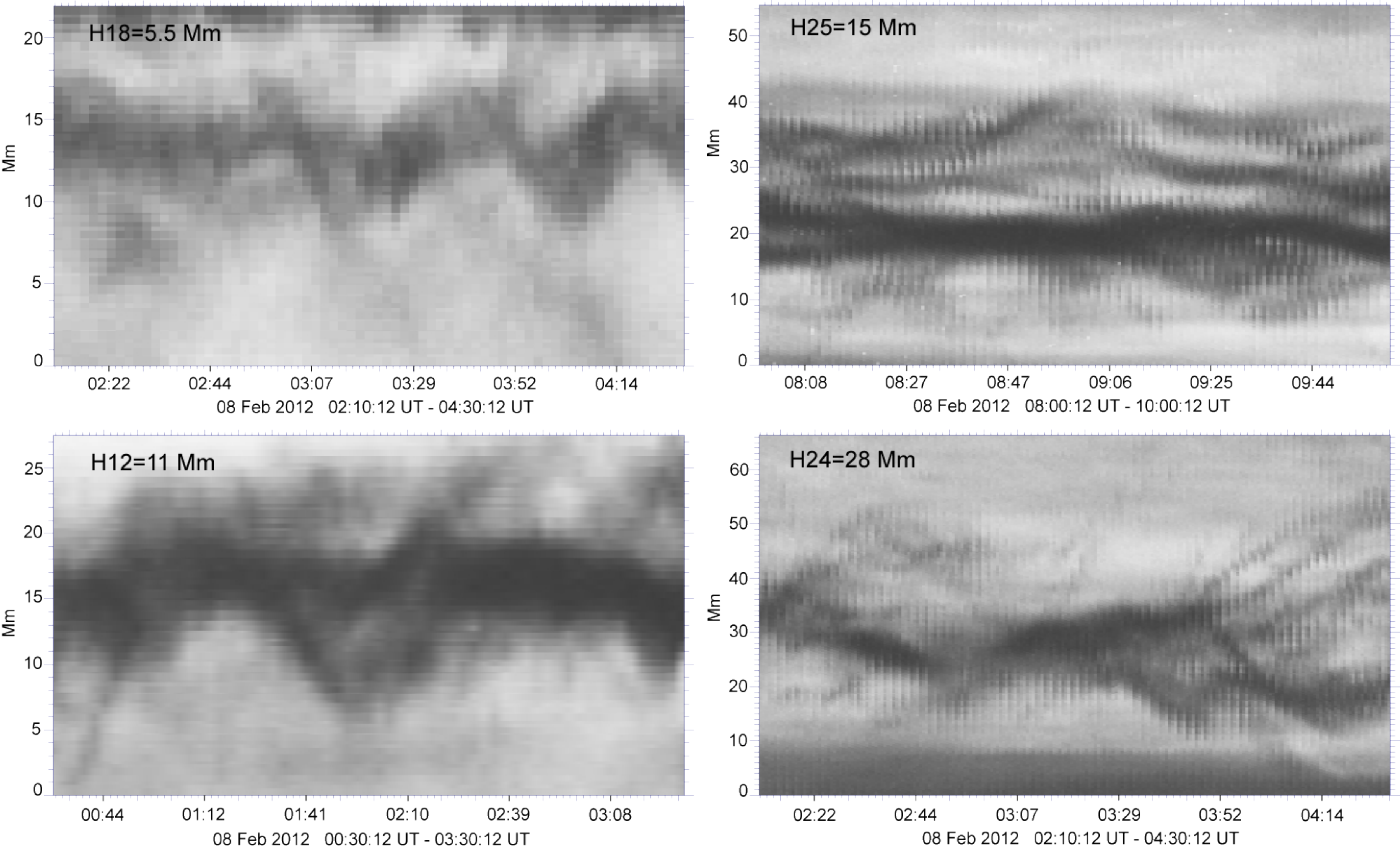}
\caption{Time-slice diagrams for cuts at different heights (H18, H12, H25, H24 in ascending order) in prominence barb B1 from Figure~5 taken by SDO/AIA in the 171\,\AA\ spectral line with two-minute cadence on 8 February 2012. The overall duration is 120 minutes for panel H25,  140 minutes for panels H18 and H24, 180 minutes for panel H12. Oscillation periods vary from 40 minutes for H18 to 65 minutes for H12.}
\end{figure}

\section{Apparent Vortical Motions in Prominence Spines and Barbs}
   
What produces the illusion of vortical motion in apparent tornado-like prominences? It is mainly the fact that we can see plasma appearing and disappearing on the upper sides of a vertical column, which looks wider at the top. This illusion is easily produced by counterstreaming flows: indeed, the apparent rotational motion is only observed in 2D projection on the limb in the plane of the sky and never on the disk; the constant counterstreaming motion of the prominence plasma along the thin threads, especially when they connect the vertical parts of the prominence (legs and barbs) to the much more horizontal spine, creates an effect that the eye associates with rotation. But the overall prominence topology and magnetic structure is very important to understanding the true nature of such observed motions at the solar limb (Panasenco and Martin, 2008; Lin {\it et al}., 2008; Martin {\it et al}., 2012; Liewer, Panasenco, and Hall, 2013). 
   
Limb observations of a group of four tornado-like prominences in the 171\,\AA\ spectral line taken by TRACE on 27 November 1999 are shown in Figure 1 together with the position of these prominences on the limb in a full disk image. Figure 3 shows a sequence of full disk H$\alpha$ images of this group from the Big Bear Solar Observatory (BBSO) as the group rotates onto the disk. Even these low-resolution images, taken at relatively low cadence, establish a direct connection between apparent tornado-like prominences on the limb and filament structures known as barbs (Martin, 1990; 1998).  Suppose the assumption that tornadoes at the limb are the result of a vortex-like motion around a conical structure: the top view of this cone would then be described by a circle with diameter equal to the width of the cone top as viewed from the side, {\it i.e.} on the limb. Measurements show that the width of the tornado-like cone near the top, where the strongest motion occurs, is $\approx$ 100 Mm as observed by TRACE (Figure 1). However, the top view of the same structure seen on the disk has a width of $\approx$~5\,--\,10 Mm, corresponding to the filament spine width (Figure 3). The vortical interpretation therefore cannot stand, and we conclude that observed illusion of the 3D rotation comes from the mostly 2D counterstreaming motion along the prominence/filament spine and barb threads. The four tornado-like prominences observed at the limb by TRACE on 27 November 1999 are simply four filament barbs on the two sides of a filament spine that had a length of $\approx$ 200 Mm on 2 December 1999 (Figure 3).

To further support these conclusions based on TRACE and BBSO 1999 data we present here two examples of more recent high-resolution and high-cadence observations obtained by the SDO/AIA instrument in February and March 2012. Figure 4 shows an apparent tornado-like prominence on the limb and its subsequent appearance on the disk. The width of the tornado observed at the limb is $\approx$ 40 Mm on 27 March, the width of the same area viewed from the top against the disk is $\approx$ 2\,--\,3 Mm on 30 March. The difference between these two measurements is an order of magnitude, substantiating the interpretation of the apparent tornado-like structure as 2D counterstreaming. An observation supporting the narrow width of the filament channel and spine in the corona comes from coronal cells, cellular features in Fe {\sc xii} 193\,\AA\ images of the 1.2 MK corona first observed and modeled by Sheeley and Warren (2012). Panasenco et al. (2012) found that coronal cells do not cross the polarity reversal boundary within a filament channel at heights below the filament spine top. Coronal cells originate from the network field concentrations and show the same pattern of chromospheric fibrils (which align along the filament channel axis) because they follow the same filament channel magnetic topology, with a (presumably) strong horizontal component of the field. The distance between coronal cells on opposite sides of the filament channel is very narrow and does not exceed 1\,--\,15 Mm, even though coronal cells reach up to 100 Mm and more in height. These coronal observations support our chromospheric measurements of the filament channels, but much higher up into the corona.

\begin{table}
\caption{ Period [T] of oscillations versus height [H] of the cuts  across Barb 1 and Barb 2 on 8 February 2012.
}
\label{T1}
\begin{tabular}{c|cccccccccc|cc}                              
  \hline            
 & \multicolumn{10}{|c|}{\bf{Barb 1}}  &   \multicolumn{2}{|c|}{\bf{Barb 2} } \\
   \hline                  
H &5&5.5&11&14&15&21&21.5&24.5&28&30&7.5&15  \\
Mm &&&&&&&&&&&\\
  \hline
  T &70&40&65&66\,--\,80&50\,--\,55&70\.--\,90&45\,--\,66&60\,--\,66&40\,--\,50&66\,--\,70&25\,--\,40&35\,--\,45 \\
min &&&&&&&&&&& \\
   \hline     
  \end{tabular} 
\end{table}

\begin{figure}
\center
\includegraphics[scale=.23]{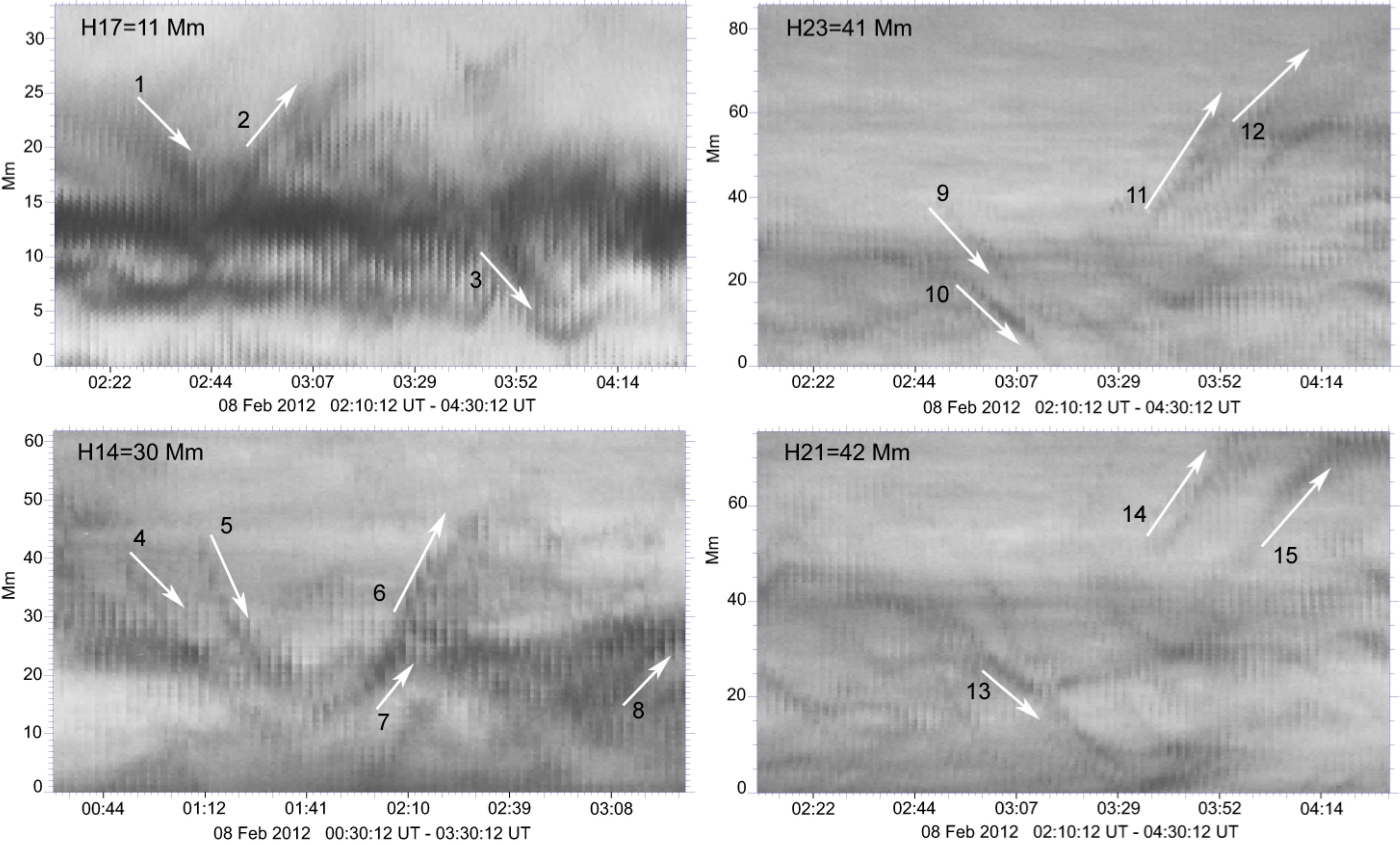}
\caption{ Time-slice diagrams for cuts at different heights (H17, H14, H23, H21 in ascending order) in prominence barb B1 from Figure 4 taken by SDO/AIA in the 171\,\AA\ spectral line with two-minute cadence on 8 February 2012. The overall duration is 140 minutes for panels H17, H23, H21 and 180 minutes for panel H14. White arrows outline counterstreaming signatures along the slices and corresponding velocities: $V_1$ = -6 $\pm$ 2 km s$^{-1}$; $V_2$ = 10 $\pm$ 2 km s$^{-1}$; $V_3$ = -10 $\pm$ 2 km s$^{-1}$; $V_4$ = -17 $\pm$ 2 km s$^{-1}$; $V_5$ = -27 $\pm$ 3 km s$^{-1}$; $V_6$ = 27 $\pm$ 3 km s$^{-1}$; $V_7$ = 15 $\pm$ 2 km s$^{-1}$; $V_8$ = 13 $\pm$ 2 km s$^{-1}$; $V_9$ = -21 $\pm$ 3 km s$^{-1}$; $V_{10}$ = -16 $\pm$ 2 km s$^{-1}$; $V_{11}$ = 26 $\pm$ 3 km s$^{-1}$; $V_{12}$ = 14 $\pm$ 2 km s$^{-1}$; $V_{13}$ = -14 $\pm$ 2 km s$^{-1}$; $V_{14}$ = 25 $\pm$ 3 km s$^{-1}$; $V_{15}$ = 21 $\pm$ 3 km s$^{-1}$.}
\end{figure}

\begin{figure}
\center
\includegraphics[scale=.35]{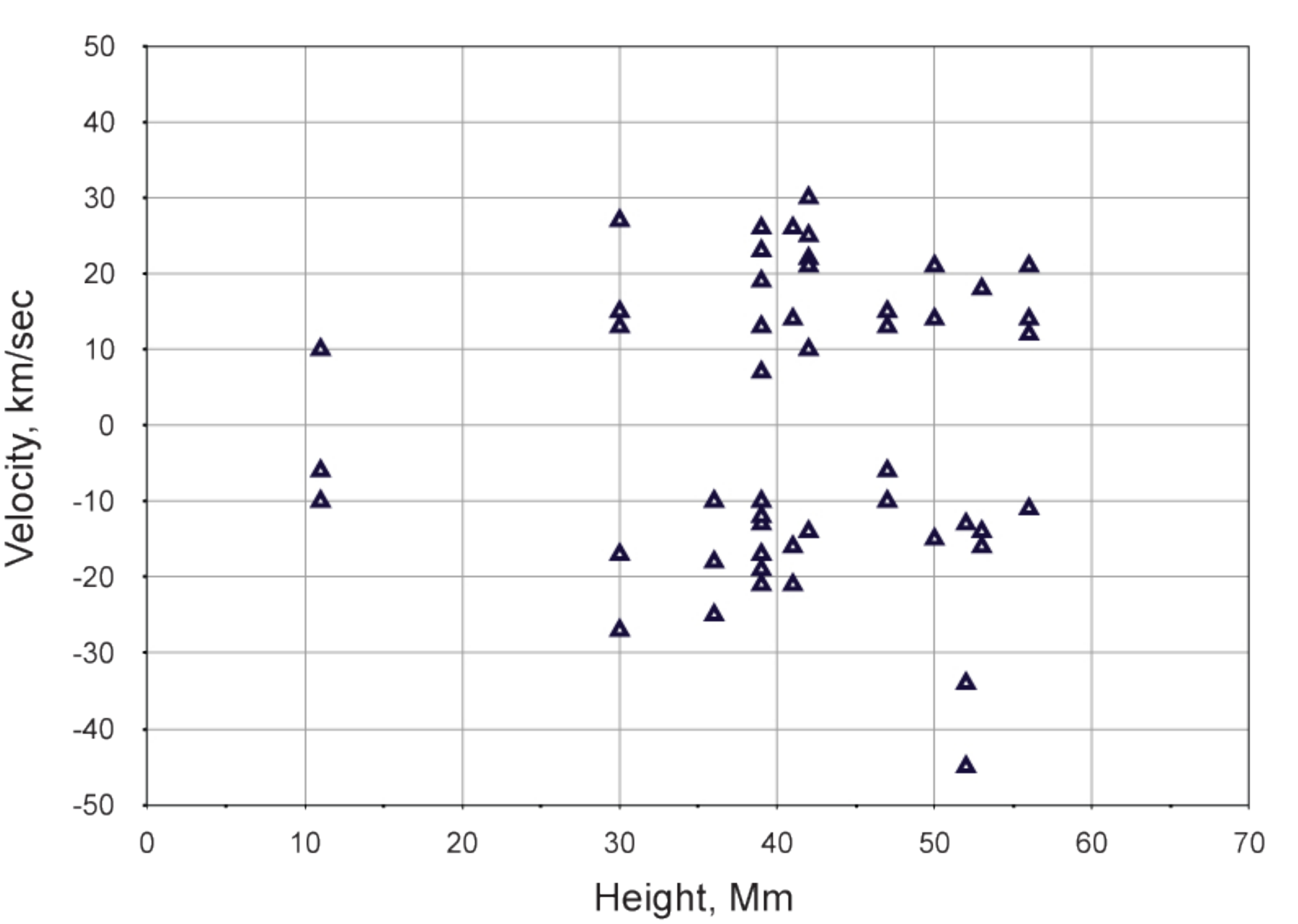}
\caption{ Distribution of counterstreaming velocities measured along cuts at different heights.}
\end{figure}

\begin{figure}
\center
\includegraphics[scale=.19]{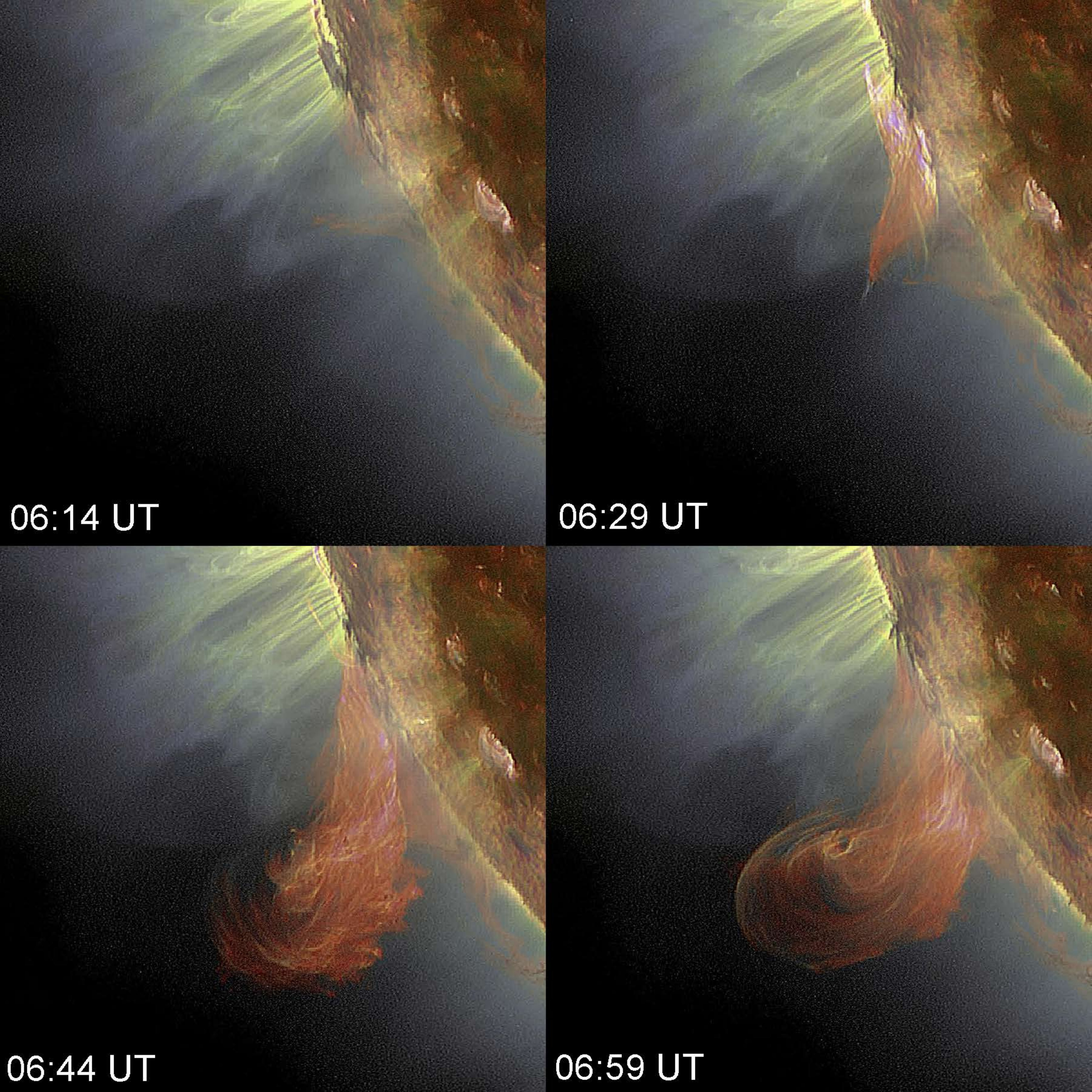}
\caption{ The plasma motion along the filament cavity when the filament channel is crossing the limb on 16 June 2011. Superposition of 171\,\AA, 193\,\AA, and 304\,\AA\ spectral lines, SDO/AIA.}
\end{figure}

The speeds of counterstreaming plasma motions in apparent tornado-like prominences seen on the limb (along the prominence threads observed in the 304\,\AA\ spectral line on 27 March 2012) was estimated to be up to $45\pm5$\,km s$^{-1}$ in the plane of the sky. Oppositely directed plasma motion along the threads at different heights contributes to the illusion of rotational tornado-like motion. A similar apparent tornado-like prominence was observed on 7\,--\,8 February  2012 (see supplementary movies), but it erupted on 9 February, so we were not able to trace it to the disk as we did for the 27 March prominence. The region of interest of the filament channel was nearly parallel to the limb, allowing us to make measurements of the plasma speeds along the prominence spine and across barbs in more preferable observational conditions.  Figure 5 shows the prominence with barbs observed on the limb in the 171\,\AA\ spectral line by SDO on 8 February 2012. We used 28 cuts to create time-slice diagrams for the lower barb regions, upper barb regions and prominence spine. The time-slice diagrams for the low barb regions are shown in Figures 6 and 7. Because barbs connect the prominence spine in the corona with the photosphere the plasma motion along these lower part of the barbs is mostly vertical and we do not observe continuous horizontal motions here but sporadic ejections of plasma. However, it is interesting to notice that the trunks of the prominence barbs show oscillations with an average period $\approx$ 40\,--\,70 minutes. These oscillations could also cause an illusion of rotational motion. Figure 6 shows two time-slice diagrams for cuts H2 and H6 in prominence barb B1 (see Figure 5) taken at heights 21.5 Mm and 30 Mm respectively. The right panel of this figure shows the same time-slice diagrams with added white lines showing barb sub-structure as an aid to illustrate the calculation of oscillation periods, ranging from 45 to 70 minutes for the sub-structures. Figure 7 shows four time-slice diagrams for cuts at different heights (H18, H12, H25, H24 in ascending order) in prominence barb B1, with oscillation periods varying from 40 minutes for H18 to 65 minutes for H12. Table 1 summarizes all observed oscillations in the low parts of the prominence barbs. Filament spine and barbs are composed of thin threads. The widths of filament threads in H$\alpha$ are $\le$ 200 km (Lin {\it et al}., 2005a).  The oscillations in the individual H$\alpha$ threads have been found to be $\approx$ 10\,--\,20 minutes (Lin, 2011; Lin {\it et al}., 2007; 2009), but the oscillation of groups of prominence threads have been found to be $\approx$ 40 minutes (Berger {\it et al}., 2008; Panasenco and Velli, 2009). The time-slice diagrams in Figure 6 and 7 show that barbs have a substructure and are often composed of two to five bundles of threads resolved in the 171\,\AA\ spectral line. Each such bundle has an oscillation period of $\approx$ 40\,--\,70 minutes.  We will review and discuss the possible sources of these oscillations in the Discussion section.

The time-slice diagram in the bottom right panel in Figure 7 shows plasma motion along cut H24 at a height of 28 Mm. One can see that approximately at 03:52 UT one of the relatively solid bundles of the barb B1 splits into many thin separate sub-bundles where plasma is moving with a speed 10\,$\pm$\,2~km~s$^{-1}$ in opposite directions. A similar situation has been observed for cuts H2, H3, H5, and H6 with corresponding heights 21.5, 24.5, 21, and 30 Mm. This allows us to conclude that at heights above $\approx$ 21 Mm the plasma motion along the vertical part of the barbs became more and more horizontal and parallel to the solar surface. Consequently, the horizontal component of the plasma speed gradually increases from 5 km s$^{-1}$ at a height of 21 Mm up to 45 km s$^{-1}$ at a height of 54 Mm. 

The next step in our study is to measure the plasma speed along the upper barbs and spine of the prominence on 8 February 2012. These cuts correspond to areas where the plasma motion is less vertical and can be nicely estimated from time-slice diagrams. We made 14 different cuts across the upper part of the barb B1, and along the prominence spine between the barbs B1 and B2 (see labels on Figure 5). The width of the each cut is three pixels $\approx$ 1.3 Mm. Such width can accommodate at most $\approx$ 6\,--\,7 thin filament threads along which plasma is moving. The observations from SDO/AIA have a resolution which does not allow us to resolve all threads. We will estimate in our measurements the speed of plasma pieces moving in and out of our cuts at different heights. 

The time-slice diagram in the upper left panel in Figure 8 shows an exceptional horizontal motion of plasma from the barb B1 at a low height of 11 Mm along cut H17. The sporadic ejections of plasma from the barbs along the low threads happen relatively rarely, and usually have a low horizontal component of the speed, which is 6\,--\,10 km s$^{-1}$ in opposite directions in this case. The other three panels in Figure 8 correspond to heights of 30, 41, and 42 Mm. Time-slice diagrams for these heights show a wide spectrum of plasma speeds in the plane of the sky ranging from 13 to 27 km s$^{-1}$. The distribution of the plasma speed along cuts at different heights is shown in Figure 9. Here we combine all measurements along all 14 cuts across the upper part of the barb B1, along the prominence spine above the barb B1, and between barbs B1 and B2. One can see that the plasma moves in opposite directions with speeds
that have a similar distribution. These oppositely directed, mostly 2D motions at different heights create a perfect illusion of rotating vortices. The counterstreaming is easily observed in Doppler and might be interpreted, incorrectly, as a rotational motion when observed on the limb (Orozco Su{\'a}rez {\it et al}., 2012). Note that Orozco Su{\'a}rez {\it et al}. (2012) estimated this rotational velocity from Doppler measurements as $\pm$\,2\,--\,6 km s$^{-1}$, which is within the noise range for counterstreaming plasma motions along prominence threads. The filament channel orientation relative to the limb plays a crucial role in Doppler measurements. Any deviation from the parallel orientation to more perpendicular with respect to the limb  will increase the Doppler velocities due to the same counterstreaming motions. 
    
\begin{figure}
\center
\includegraphics[scale=.24]{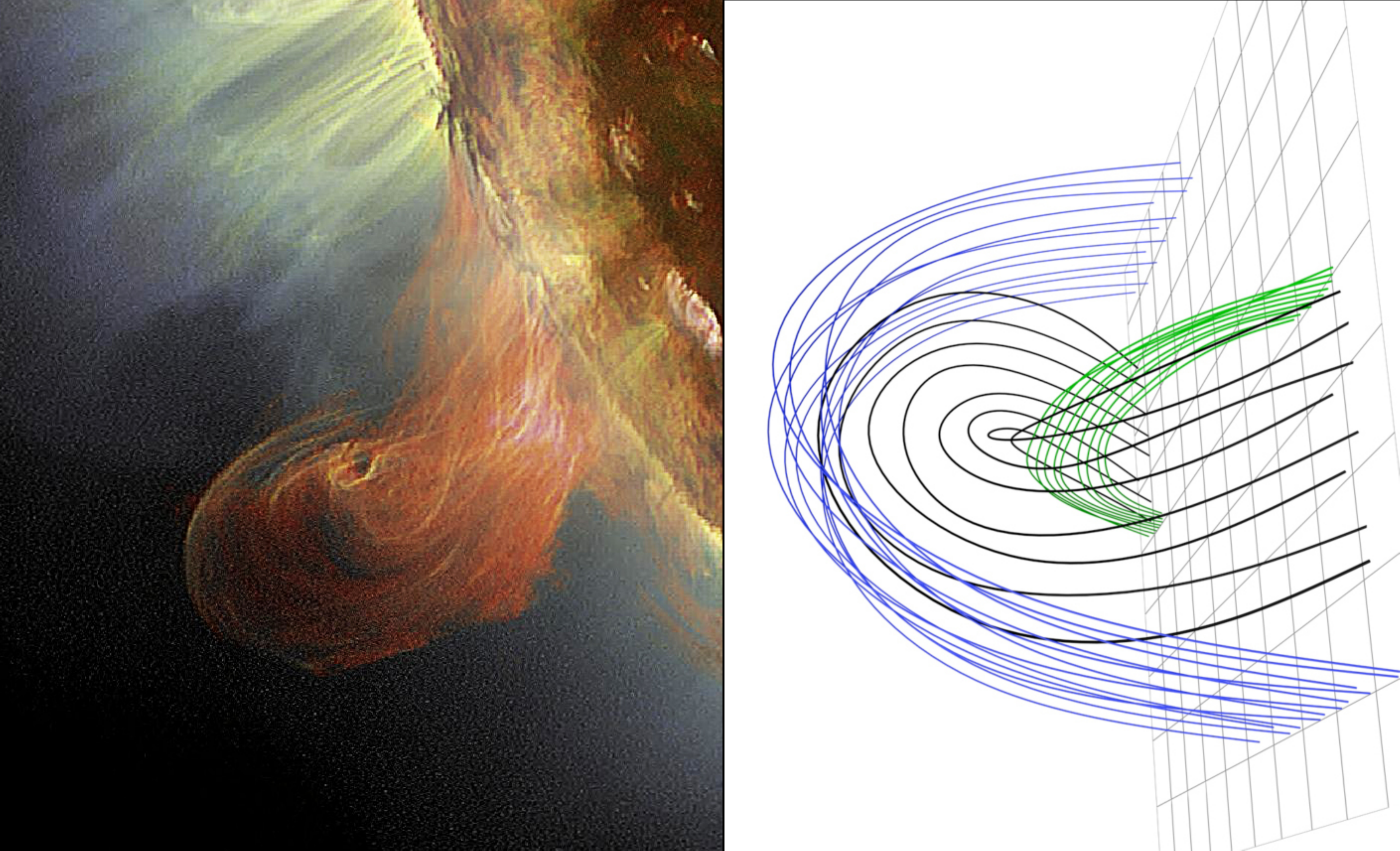}
\caption{ The similarity between the observed trajectory of the plasma moving along the filament cavity (16 June 2011, 06:59 UT, 171\,\AA, 193\,\AA, and 304\,\AA\ spectral lines, SDO/AIA) and modeled cavity with a strong writhe (black lines in the right panel) when viewed mostly along the filament channel. }
\end{figure}

\section{Apparent Rotational Motions in Prominence Cavities Due to Writhe}

Apparent tornado-like motions on the limb can also be observed when rapidly moving plasma from an outside source propagates into and along a filament channel cavity. In this case the plasma motion does have a 3D geometry, but it does not develop along the ribbon-like filament spine or barbs, rather along the filament channel cavity magnetic field lines, which have a strong writhe (black lines in Figure 2). Usually the source of this plasma is from a neighboring active region. For quiescent filaments, the injections of plasma inside the filament cavity occur during flare-like activity. Sporadic injections of plasma inside the filament channel cavity create an illusion of rotational plasma motion on the limb, when observed along the filament channel axis as the filament channel rotates through the limb. This illusion is created by the projection of the writhed magnetic field lines onto the limb plane. The mystery is easily resolved if limb observations from L1 or the Earth point of view are compared with simultaneous observations by the {\it Sun-Earth Connection Coronal and Heliospheric Investigation} (SECCHI: Howard {\it et al}., 2008) onboard the {\it Solar TErrestial RElations Observatory} (STEREO: Kaiser {\it et al}., 2008).
    
Figure 10 shows multispectral observations from SDO/AIA of the plasma motion with speed up to 200 km s$^{-1}$ along the filament cavity when the filament was crossing the limb on 16 June 2011. The speed was measured using the 195\,\AA\ spectral line STEREO-B images of this plasma motion when observed against the disk. The separation angle between STEREO-B and SDO for 16 June was $\approx$ 93 degrees. The position of the active region -- the source of sporadic plasma injections inside the filament cavity -- is S16W7 for STEREO-B, and exactly at the limb for SDO. Figure 11 shows the similarity between the observed trajectory of the plasma moving along the filament cavity (16 June 2011, 06:59 UT, SDO/AIA) and a simple cavity model including fields with a strong writhe (black lines in the right panel) when viewed mostly along the filament channel.  The black lines of the modeled cavity show the same loop-like geometry in the upper regions, close to the overlying coronal arcade (blue lines), clearly a projection effect of the lines with writhe.

\begin{figure}
\center
\includegraphics[scale=.137]{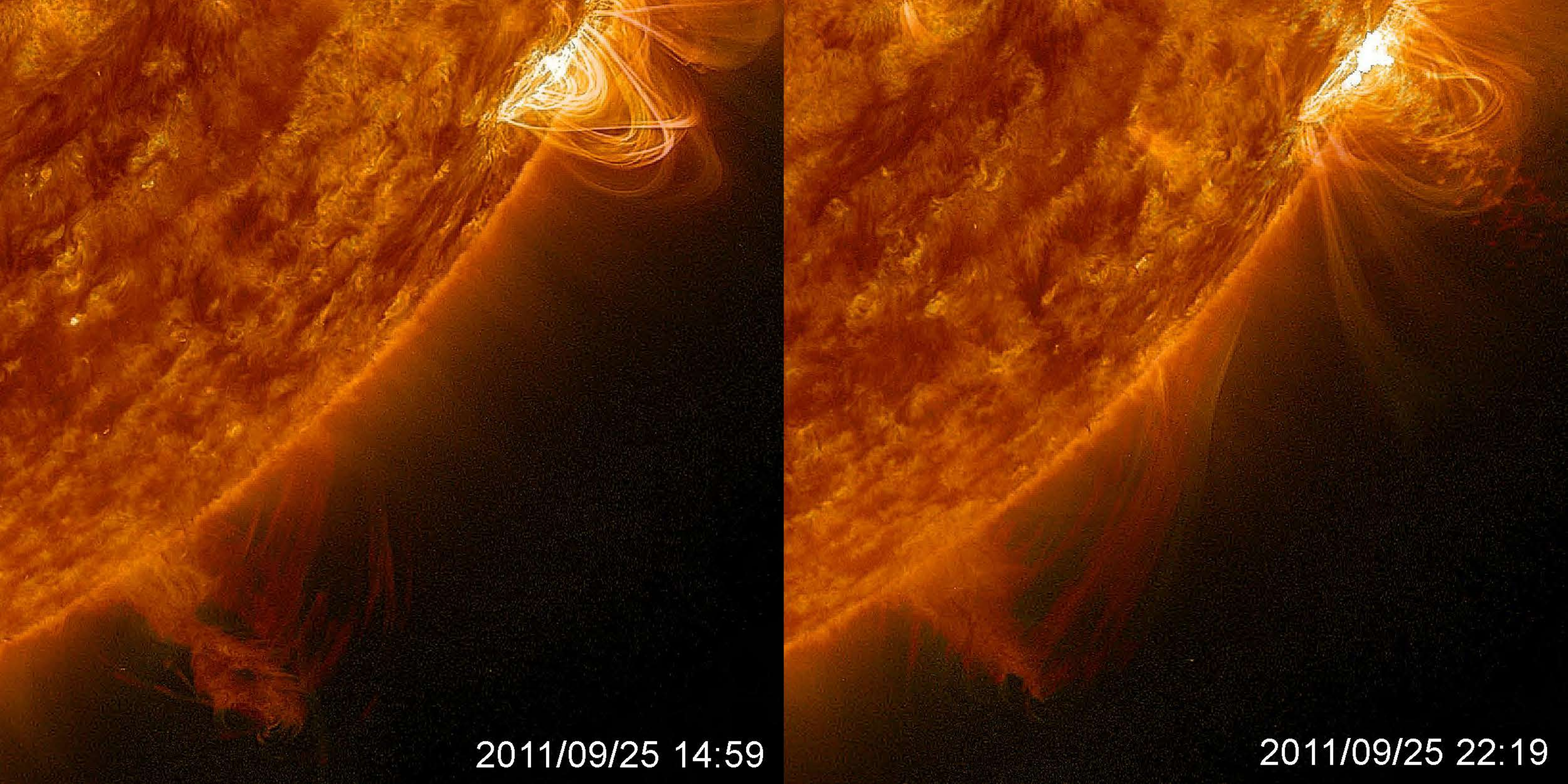}
\caption{ The plasma motion along the prominence spine and cavity from the eastern  to western prominence footpoints after the flare of class M1.5 in the active region NOAA 11303 (09:25), which was located  $\approx$ 20$^\circ$ northward from the eastern end of the polar crown filament.}
\end{figure}

\begin{figure}
\center
\includegraphics[scale=.28]{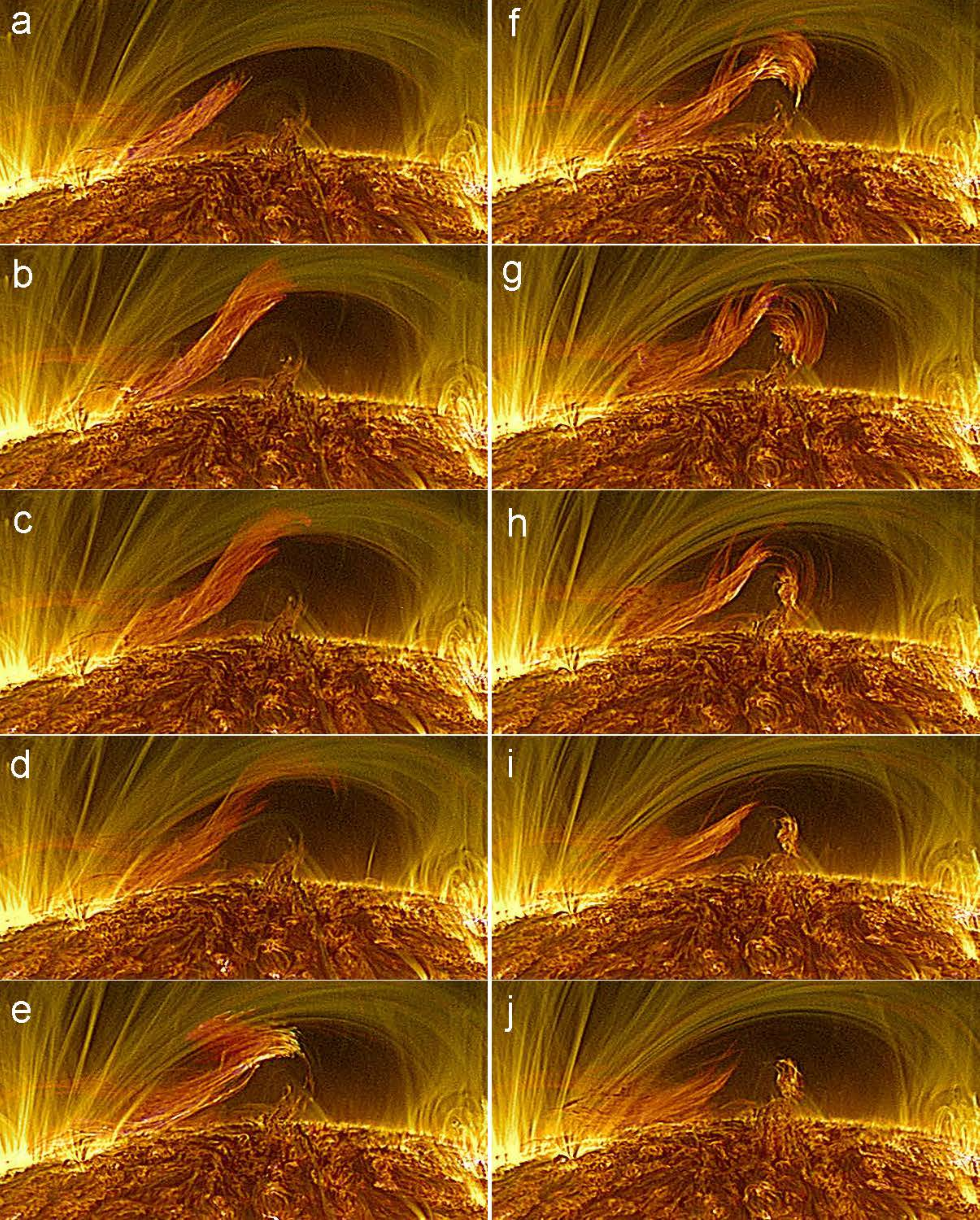}
\caption{ Plasma motion inside the filament cavity for the case when the filament is mostly parallel to the limb. Superposition of 171\,\AA and 304\,\AA spectral lines, SDO/AIA (images rotated 90$^\circ$ clockwise). The time sequence of images is from 06:15 UT to 08:16 UT on 23 July 2012.}
\end{figure}

Another writhing motion in a prominence cavity was observed on 25 September 2011 and has been described in detail by Li {\it et al}. (2012) and Panesar {\it et al}. (2013). Figure 12 shows the eastern end of a polar crown filament as observed at the limb on 25 September 2011. Two bundles of thin threads were visible on 24 September -- two barbs, one visible on the disk and at the limb, and another with its footpoint completely behind the limb. With solar rotation, these barbs co-align and superpose, so only one barb can be observed on 25 September. The active region NOAA 11303, located $\approx$ 20 degrees northward from the eastern filament end, produced multiple flares. The strongest flare (class M1.5) occurred on 25 September, 09:25 UT. This last flare created coronal disturbances with subsequent strong oscillations of the whole prominence system, triggering magnetic reconnection of the fields in the prominence barbs and footpoints.  As a result of such reconnections, plasma was injected inside the prominence system from the east end of the channel, and later propagated along the prominence spine and cavity in opposite directions toward both filament footpoints (Figure 12). The illusion of a tornado was created by the projection effect together with the oscillatory disturbances caused by the active region flaring only 20 heliographic degrees from the filament.  Some possible reasons for such oscillatory disturbance are described by Panesar {\it et al}. (2013) and include the Hudson effect (Hudson, 2000).
    
When the position of the filament channel is nearly parallel to the limb with a very small acute angle, the plasma injected inside the filament cavity produces a spectacular picture as it traces the magnetic topology of the cavity (where the plasma beta is arguably less than unity). The observed lines exhibit clear writhe. Figure 13 shows a sequence of images of the plasma moving inside the filament cavity on 23 July 2012 from 06:15 to 08:16 UT. Panels (g) and (h) in Figure 13 show the cavity maximally filled with plasma tracing the field, which can be best described as bundles with writhe.  
    
The sporadic propagation of the externally injected plasma inside the filament cavity is not a rare phenomenon. It is a very important tool in our understanding of the magnetic topology of the filament system, which also includes the filament channel cavity field lines. These lines usually are not traceable due to the very low plasma density of the cavity compared with the filament/prominence or surrounding corona. The sporadic injections of the plasma inside the filament channel system work as perfect tracers to uncover the cavity magnetic topology.

\begin{figure}
\center
\includegraphics[scale=.24]{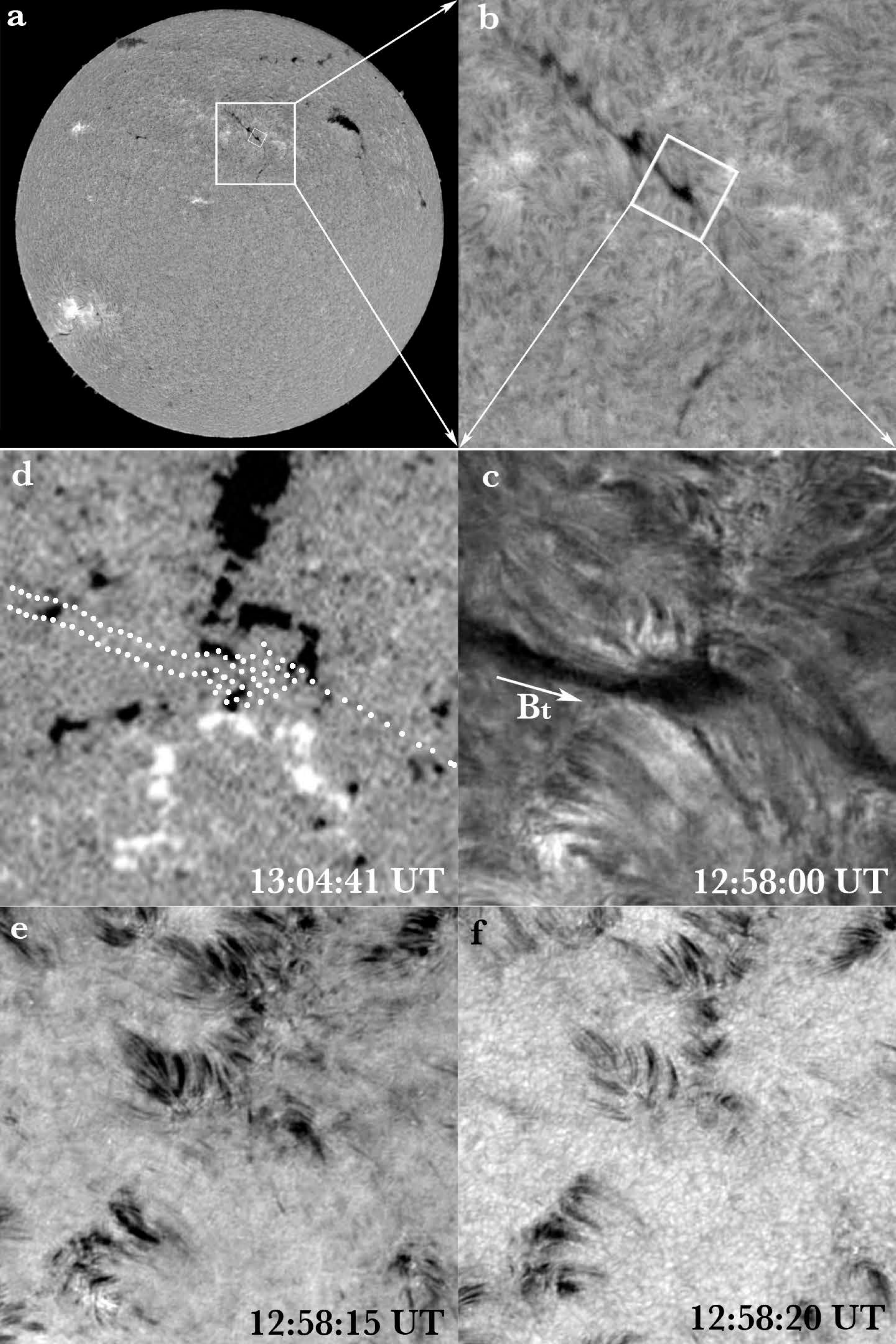}
\caption{The magnetic boundary at the source of barb. a) a full disk  H$\alpha$ image taken by the BBSO on September 13, 2010. The white square is centered around a dextral filament which is displayed in the panel b); b) the white square shows part of the filament as observed by the Dutch Open Telescope in the core of the H$\alpha$ line and shown, rotated, in the panel c). The direction of the horizontal component of the magnetic field $B_t$ along this dextral filament channel is given by the white arrow.  Panel d) is the corresponding magnetogram from SDO/HMI. The dotted line superimposed on the magnetogram outlines the filament barb anchored at the junction of four supergranular cells. Panels d) and f) are images in the blue and red wings of (H$\alpha$ -/+0.07nm) respectively. The field of view spans 45.5 x 45.5 Mm for panels c-f.}
\end{figure}

\begin{figure}
\center
\includegraphics[scale=.19]{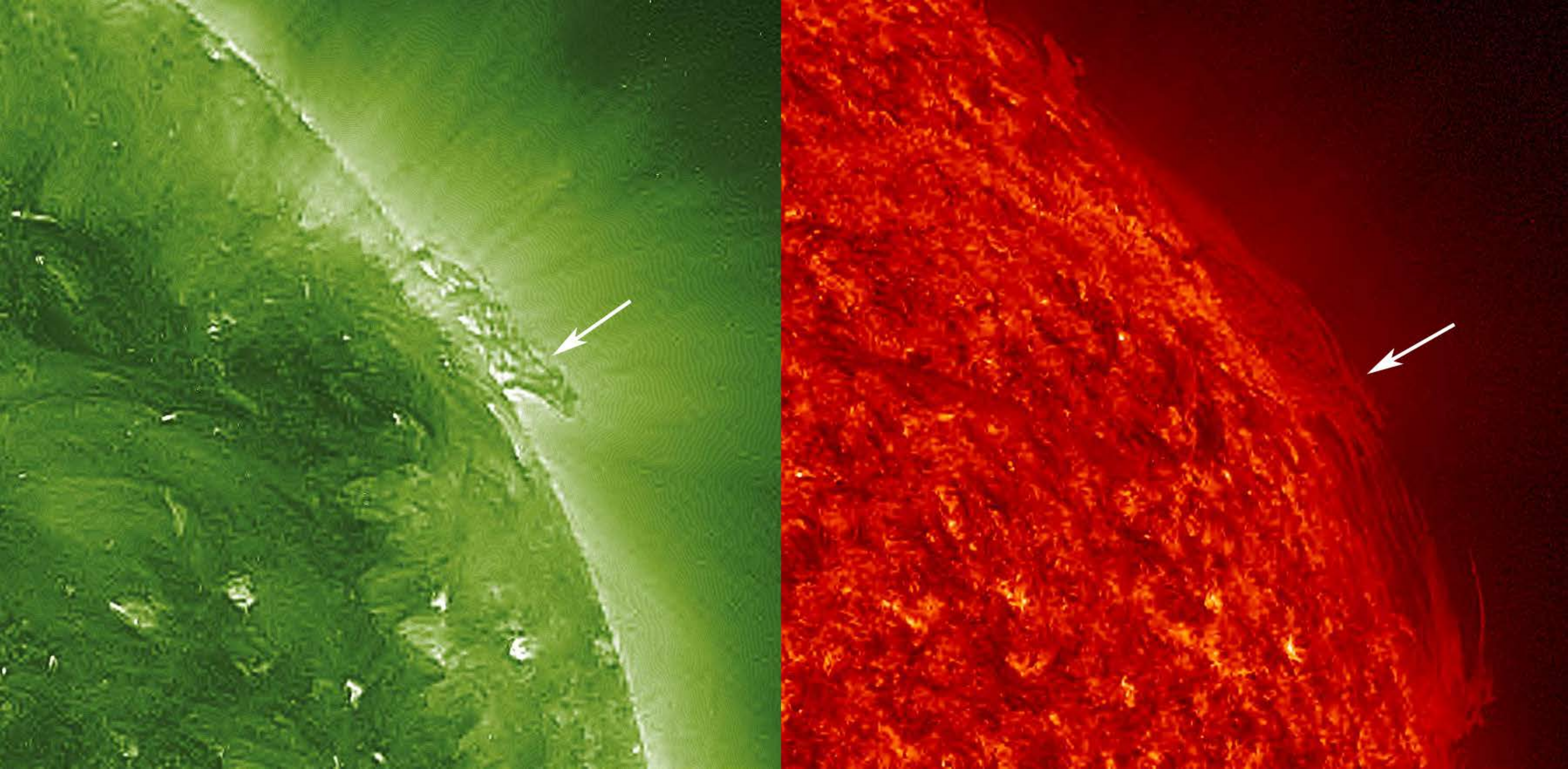}
\caption{ "Tornado prominences" observed by SDO/AIA 193\AA\  at the west limb on September 20, 2010 (left) and its appearance in 304\AA\ as a regular prominence (right). One of these "tornados", pointed with a white arrow, corresponds to the barb of the filament observed at the disk on September 13 (Figure 14).}
\end{figure}

\section{Discussion} 

Lin {\it et al}. (2005b) have shown that close to 65\,\%\ of the observed end points of barbs fall within photospheric network boundaries. Recent high-resolution observations at the {\it Dutch Open Telescope} (DOT) reveal that filament barbs are anchored at the intersections of supergranular cells (Figure 14). We trace the filament observed by Big Bear Solar Observatory and the part of its spine and barb observed by DOT on 13 September 2010 to the West limb, where the corresponding prominence was observed by SDO/AIA in 304\,\AA\ (Figure 15).  When observed in 171\,\AA\ this prominence turns out to be a group of apparent tornado-like prominences. One is the barb observed in high resolution by the DOT. The fact that the barb is rooted in the junction of four supergranular cells creates conditions for magnetic field canceling (Litvinenko, 1999, 2010), a necessary requisite for filament formation, since supergranular intersections are perfect locations where opposite polarities can be pushed together and cancel.  Ii is interesting to note that the DOT observations in the far-blue and -red wings (shown in panels (e) and (f) in Figure 14) demonstrate the pattern of fibrils in the filament channel. They are nearly parallel to the filament spine and do not cross it. The same pattern was observed in coronal cells (Panasenco {\it et al}., 2012), and provides additional observational evidence of how narrow a filament spine at the chromospheric level is -- the distance between two oppositely directed fibrils being  $\approx$ 4.5\,--\,6 Mm (Figure 14 (e), (f)) --the same order as the width of  filament spines (2\,--\,3 Mm) and the distance between oppositely directed coronal cells (10\,--\,15 Mm).

Supergranular cell intersections are also natural locations where photospheric convective motions turn into downdrafts, and downdrafts are also locations where fluid vortical motions can develop (Attie {\it et al}., 2009). A number of recent articles have linked the apparent vortical motions seen in the outer solar atmosphere, at different heights in the solar chromosphere or corona, to vortices coming from the unwinding of magnetic fields entrained by convection in the much higher density photosphere (Su {\it et al}., 2012; Wedemeyer-B{\"o}hm {\it et al}., 2012). That photospheric vortices might be a source of the large amplitude Alfv\'enic turbulence observed in the solar wind had already been suggested by Velli and Liewer (1999). Indeed, photospheric supergranulation flows concentrate magnetic flux at the network boundaries, where the resulting flux tubes exceed dynamic pressure equipartition and are close to evacuation pressure-balance. Vorticity obeys the same equation as magnetic induction, so in regions where the field does not dominate the dynamics, vorticity and magnetic flux will concentrate in the same regions of space. Vorticity filaments are the natural dissipative structures of 3D hydrodynamic turbulence and are observed to form in simulations of the solar convection zone (see, {\it e.g}., Brandenburg {\it et al}., 1996). 

Simon and Weiss (1997) gave several examples of vorticity sinks, associated with photospheric downdrafts, at mesogranular scales. They fit the observed vorticity with a profile $$\omega(r)= \bigl(V/R\bigr)\ {\rm exp}(-r^2/R^2),$$
where $V$ is the characteristic rotational velocity
associated with the vortices, and $R$ their characteristic
radius. A typical photospheric vortex lasts several hours and takes
about two hours to develop. The strongest vortex that they observed
had $|\omega|\approx 1.4 \ 10^{-3}$ rads s$^{-1}$, a size $R\approx 2.5 \ 10^3$ km,
and a maximum azimuthal velocity of order 0.5 km s$^{-1}$. Coincidentally, Su et al. (2012) associate the apparent tornado-like motion of barbs with photospheric vortices of approximately the same dimensions and vorticity. 
A vortex of this type is associated with periods of hours, and not tens of minutes as the barb oscillations, seen by Su {\it et al}. (2012) and in this article, display.

Photospheric vortices on open field lines would produce an\,Alfv\'en wave\,packet, as the magnetic field lines are entrained by the rotational motion,  whose frequency is given by the vorticity itself, and duration coincides with the life of the vortex. Such Alfv\'enic-like motions would propagate upwards into the solar wind, evolving due to nonlinear interactions in the process, and amplifying and decaying with height over several solar radii. On closed field lines, such as barbs however, the twist injected by a photospheric vortex would lead to the formation of tangential discontinuities: it is fundamental to remark here that photospheric tornadoes with periods of hours would, at coronal heights, appear as quasi-steady motions (the travel time along a barb over a distance  of say, 50\,000 km would require only a fraction of a minute for a coronal Alfv\'en speed of 1000 km s$^{-1}$). Rappazzo {\it et al}. (2013)  have studied the evolution of a coronal volume subject to such vortical forcing at the photosphere and showed that the resulting field lines would first develop twist and then become unstable to an internal kink mode, releasing most of the stored magnetic energy and removing twist from the field lines, over periods of tens of Alfv\'en crossing times ({\it i.e.} tens of minutes). This being the case, it is difficult to interpret barb motions as tornadoes: a more realistic suggestion, given that barb strands are indeed formed at intersections of supergranules, is that reconnection releasing the twisted magnetic field component generated from the vortices produce mass motions along the barb itself which leads to the observed oscillations. In other words, though the photospheric vortical motions might contribute to barb formation and dynamics, this would more likely be indirectly, through the dissipation and reconnection of the vortical magnetic field.

A reliable estimate of the contribution of energy in photospheric vorticity sinks to coronal dynamics and heating, as channeled by the magnetic field and rotational motions, would really require a long-term statistical analysis of the distribution of the number, intensity, and duration of such vorticity tubes in the photosphere, which will hopefully become possible with the extended SDO/HMI observations. As the present discussion has shown, however, the interpretation of the motions in prominence barbs and spines as direct transmission of such vortical motions seems unlikely.

\section{Conclusions}

Observations of solar apparent tornado-like prominences have been interpreted as extraordinary vortical motions in the solar atmosphere. However, a careful analysis of older events seen on the limb and then on the disk from space and ground-based instruments, together with the multi-wavelength images from SDO and the multiple viewpoints provided by STEREO, reveal solar apparent tornado-like prominences to be due to two different type of illusions involving the motions of the plasma in the solar atmosphere as projected onto the plane of the sky.
    
We have shown that apparent vortical-like motions in prominences are really the projection above the solar limb of the mostly 2D counterstreaming plasma motion and oscillations along the prominence spine and barbs. A writhing motion, creating an illusion of a tornado, on the other hand, consists in the limb projection of the 3D plasma motion following the magnetic fields inside and along the prominence cavity. 
    
The impression of tornado-like rotational motion results in both cases from plasma motion and oscillations along magnetic field lines observed on the plane of the sky, rather than from a true vortical motion around an apparent vertical or radial axis. Apparent tornado-like prominences, once understood, provide a tool to understand the magnetic structure of filament channels, filaments, and prominences, whose role in space weather is very important.  Most coronal mass ejections (CMEs) originate from coronal loop systems surrounding filament channels. Understanding the correct magnetic structure of the filament channel system will shed light on the formation of the CMEs and their propagation in the corona.

This interpretation of apparent tornado-like prominences does not mean that rotational motions or swirls in the chromosphere and corona are absent (especially at smaller scales and times: Wedemeyer-B{\"o}hm {\it et al}., 2012). Indeed, such motions may be present and participate, as large amplitude 
Alfv\'enic motions, in the heating of the solar corona and acceleration of the solar wind along open field lines (Verdini {\it et al}., 2010). In closed field regions however, it is much more likely that such twists must relax in tangential discontinuities and current sheets as Parker nanoflares (Rappazzo {\it et al}., 2013), and in injected plasma flows into the barbs and up into the prominence spines (Cirtain {\it et al}., 2013).

\begin{acks} 
OP and SM are supported in this research under the NASA grant NNX09AG27G and NSF SHINE grant 0852249.  The work of MV was conducted at the Jet Propulsion Laboratory, California Institute of Technology under a contract from the National Aeronautics and
Space Administration. We are thankful to Aram Panasenco for contribution in image processing. The SECCHI data are produced by an international consortium of the NRL, LMSAL and NASA GSFC (USA), RAL and Univ. Bham (UK), MPS (Germany), CSL (Belgium), IOTA and IAS (France). The AIA data used here are courtesy of SDO (NASA) and the AIA consortium. The {\it Dutch Open Telescope} (DOT) is located at Observatorio del Roque de los Muchachos (ORM) on La Palma. The DOT was designed and built by Rob H. Hammerschlag. We thank the referee for interesting comments and suggestions.
\end{acks}

\end{article}
\end{document}